\documentclass{optica-article}

\journal{opticajournal} 

\articletype{Research Article}
\usepackage{textcomp}
\usepackage{lineno}
\usepackage{siunitx}
\usepackage{float}

\begin{document}

\title{Efficient, broadly-tunable source of megawatt pulses for multiphoton microscopy based on self-phase modulation in argon-filled hollow-core fiber}
\author{Yishai Eisenberg\authormark{1, *}, Wenchao Wang\authormark{1}, Shitong Zhao\authormark{1}, Eric S. Hebert\authormark{1}, Yi-Hao Chen\authormark{1},
 Dimitre G. Ouzounov\authormark{1}, Hazuki Takahashi\authormark{1}, Anna Gruzdeva\authormark{2},  Aaron K. LaViolette\authormark{1}, Moshe Labaz\authormark{3}, Pavel Sidorenko\authormark{4}, Enrique Antonio-Lopez\authormark{5}, Rodrigo Amezcua-Correa\authormark{5}, Nilay Yapici\authormark{2}, Chris Xu\authormark{1}, and Frank Wise\authormark{1}}

\address{\authormark{1}School of Applied and Engineering Physics, Cornell University, Ithaca, New York 14853, USA\\
\authormark{2}Department of Neurobiology and Behavior, Cornell University, Ithaca, New York 14853, USA\\
\authormark{3}Department of Physics and Solid-State Institute, Technion – Israel Institute of Technology, 32000 Haifa, Israel\\ 
\authormark{4}Department of Electrical and Computer Engineering and Solid-State Institute, Technion – Israel Institute of Technology, 32000 Haifa, Israel\\
\authormark{5}CREOL, The College of Optics and Photonics, University of Central Florida, Orlando, Florida 32816, USA\\}

\email{\authormark{*} ye44@cornell.edu} 


\begin{abstract*} 
An exciting recent development for deep-tissue imaging with cellular resolution is three-photon fluorescence microscopy (3PM) with excitation at long wavelengths (1300 and 1700~nm). In the last few years, long-wavelength 3PM has driven rapid progress in deep-tissue imaging beyond the depth limit of two-photon microscopy, with impacts in neuroscience, immunology, and cancer biology. However, wide adoption of 3PM faces challenges. Three-photon excitation (3PE) is naturally weaker than two-photon excitation, which places a premium on ultrashort pulses with high peak power. The inefficiency, complexity, and cost of current sources of these pulses present major barriers to the use of 3PM in typical biomedical research labs. Here, we describe a fiber-based source of femtosecond pulses with multi-megawatt peak power, tunable from 850~nm to 1700~nm. Compressed pulses from a fiber amplifier at 1030~nm are launched into an antiresonant hollow-core fiber filled with argon. By varying only the gas pressure, pulses with hundreds of nanojoules of energy and sub-100~fs duration are obtained at wavelengths between 850 and 1700~nm. This approach is a new route to an efficient, robust, and potentially low-cost source for multiphoton deep-tissue imaging. In particular, 960-nJ and 50-fs pulses are generated at 1300 nm with a conversion efficiency of 10\%. The nearly 20-MW peak power is an order of magnitude higher than the previous best from femtosecond fiber source at 1300~nm. As an example of the capabilities of the source, these pulses are used to image structure and neuronal activity in mouse brain as deep as 1.1~mm below the dura.\end{abstract*}

\section{Introduction}
Lasers based on doped silica fibers offer many advantages, such as excellent thermal properties, nearly diffraction-limited beam quality, compactness, and mechanical stability. They have found applications in various areas, such as material processing, medicine, and nonlinear microscopy \cite{Shi:FiberLaserReview, XuWise:FiberLaserMicroscopyReview}. Many of these applications benefit from a high peak power, which makes short-pulse fiber lasers especially useful.  At repetition rates of a few hundred kilohertz, fiber chirped-pulse amplifier (CPA) systems can robustly provide sub-picosecond pulses with tens or even hundreds of microjoules of energy \cite{hundred_uJ_500_kHz_CPA, rod_amplifier_50_uJ_500_kHz, fiber_CPA_review}, enabling high optical intensity.

However, fiber lasers and amplifiers suffer from two significant drawbacks. First, they have limited wavelength tunability. Easily-obtainable fibers are typically doped with ytterbium, neodymium, erbium, or thulium, which have limited emission bandwidths centered at 1030~nm, 1064~nm, 1550~nm, and 2000~nm, respectively. Thus, fiber lasers are not directly suitable for applications that require other wavelengths. Second, optical fibers have long lengths of interaction between the light and the material guiding it. This can exacerbate dispersive and nonlinear effects, which can distort a pulse of light. Since the fibers are typically silica, self-focusing precludes them from supporting pulses with peak powers higher than a few megawatts, and various other nonlinear effects can degrade the temporal profile of pulses even at much lower peak powers. Thus, pulses emitted from fiber laser systems are typically limited to more modest energies and peak powers than pulses from their solid-state counterparts.

Three-photon fluorescence microscopy enables imaging with cellular resolution deep in scattering tissue. Ultrashort pulses with peak power above 1 MW are required for efficient fluorescence excitation, and the optimal wavelengths for deep imaging are around 1300 and 1700~nm \cite{Horton_3_photon}.  The most-common sources of pulses for three-photon microscopy convert pulses from Yb-based lasers at 1050 nm to the desired wavelengths through noncollinear optical parametric amplification (NOPA) in bulk crystals. Commercial versions typically have a conversion efficiency (to 1300~nm) of 4\% or lower\cite{SpiritNOPA, OperaF}. The pulses that drive the conversion process must therefore be very energetic, which limits the repetition rates that can be achieved. More-efficient and fiber-based sources would be attractive for multiphoton microscopy, but to date fiber sources have not reached the performance of NOPA systems.  

Researchers have implemented various wavelength-conversion techniques in fiber. These include optical parametric amplification \cite{Bigourd:FiberOPA}, Raman amplification \cite{dianov2004raman}, soliton self-frequency shift \cite{Mitschke:OriginalSSFS}, and  continuum generation \cite{1970_supercontinuum, fork1983femto_SCG}, to name a few. Since these processes exploit nonlinear propagation, they can actually benefit from a long nonlinear interaction length. However, excessive nonlinear phase shifts can still lead to pulse degradation, and with solid-core fiber the peak power is limited to around 1 MW\cite{Horton_3_photon, Guoqing_MW, rishoj2019SSMC}. To enable high-power nonlinear processes in optical fiber, researchers have been turning to hollow-core fibers filled with various gases \cite{2000_SCG_Ar, husakou2001supercontinuumtheory, matsubara2007_argon_compression, travers2011ultrafast, finger_Russell_no_Raman, Ouzounov:GasSSFS, yiHao_SSFS, ZuritaMiranda:GasOPA, benabid2002_first_demonstration_of_SRS_in_HCF}. Gases generally have much lower nonlinear refractive indices than glasses, which allows scaling of pulse energy to much higher peak power, while the waveguide structure provides long nonlinear interaction length. Moreover, because the nonlinearity and dispersion a pulse experiences in gas depend upon the gas pressure--which an experimenter can vary--gas offers a level of tunability absent from solid-core fiber.

Continuum generation in solid material, gas \cite{1986_gas_supercontinuum}, or solid-core fiber \cite{first_fiber_SCG}, is a ubiquitous approach to generating light in a wide range of wavelengths. Once the continuum is generated, a spectral interval within it can be selected for use in applications. This approach has been used as early as the 1970s \cite{alfano1971picosecond_spectroscopy}, and has more recently been studied systematically as a way to provide tunable pulses for nonlinear microscopy as well as other applications \cite{Guoqing:OriginalSESS, Guoqing_pre_chirp_managed}. When the continuum is created mainly by self-phase modulation (SPM) in optical fiber, the method has been referred to as "self-phase-modulation-enabled spectral selection" (SESS)\cite{Guoqing:OriginalSESS, Guoqing_pre_chirp_managed}. SPM offers certain advantages over the other wavelength-conversion processes mentioned above: wavelength tunability without a need for phase-matching, pulses at multiple wavelengths are generated simultaneously, and the possibility of generating nearly transform-limited pulses at the desired wavelengths. When implemented with glass fiber it has provided pulses of megawatt peak power with wavelengths of 1300~nm and 1700~nm \cite{Guoqing_MW}, and SESS-generated pulses have proven to be useful in multiphoton microscopy \cite{Guoqing_microscopy_images}. However, because the mode area of single-mode fiber is limited, there is no route to scaling to significantly-higher powers in solid fiber. This motivates the extension of SESS to gas-filled hollow fiber. SPM in hollow fibers filled with atomic gas has been studied numerically assuming femtosecond input pulses at 1030~nm \cite{shi2023theoreticalSESS_gas}. Experimentally, wavelength shifts of a few tens of nanometers have been used to improve the temporal contrast of a pulse \cite{JenaExperimentalSESS}. However, there is no prior report of experimental generation of broadly-tunable pulses using this approach. 
 
Here we present a wavelength-tunable source of multi-megawatt pulses based on self-phase modulation in argon-filled hollow-core fiber. With 10-\si{\micro\joule} and 100-fs pulses at 1030~nm, we generate coherent pulses with central wavelengths between 850 nm and 1700 nm and peak powers well above 1 MW. The results include the production of 960-nJ and 50-fs (>10 MW) pulses at 1300 nm, with about 10\% conversion efficiency. The effectiveness of this source is illustrated by the use of the 1300-nm pulses to image structure and neuronal activity deep in a mouse brain using three-photon microscopy.

\section{Experimental Setup}
\begin{figure}[ht!]
\hspace*{-1cm}  
\centering\includegraphics[width=0.45\textwidth]{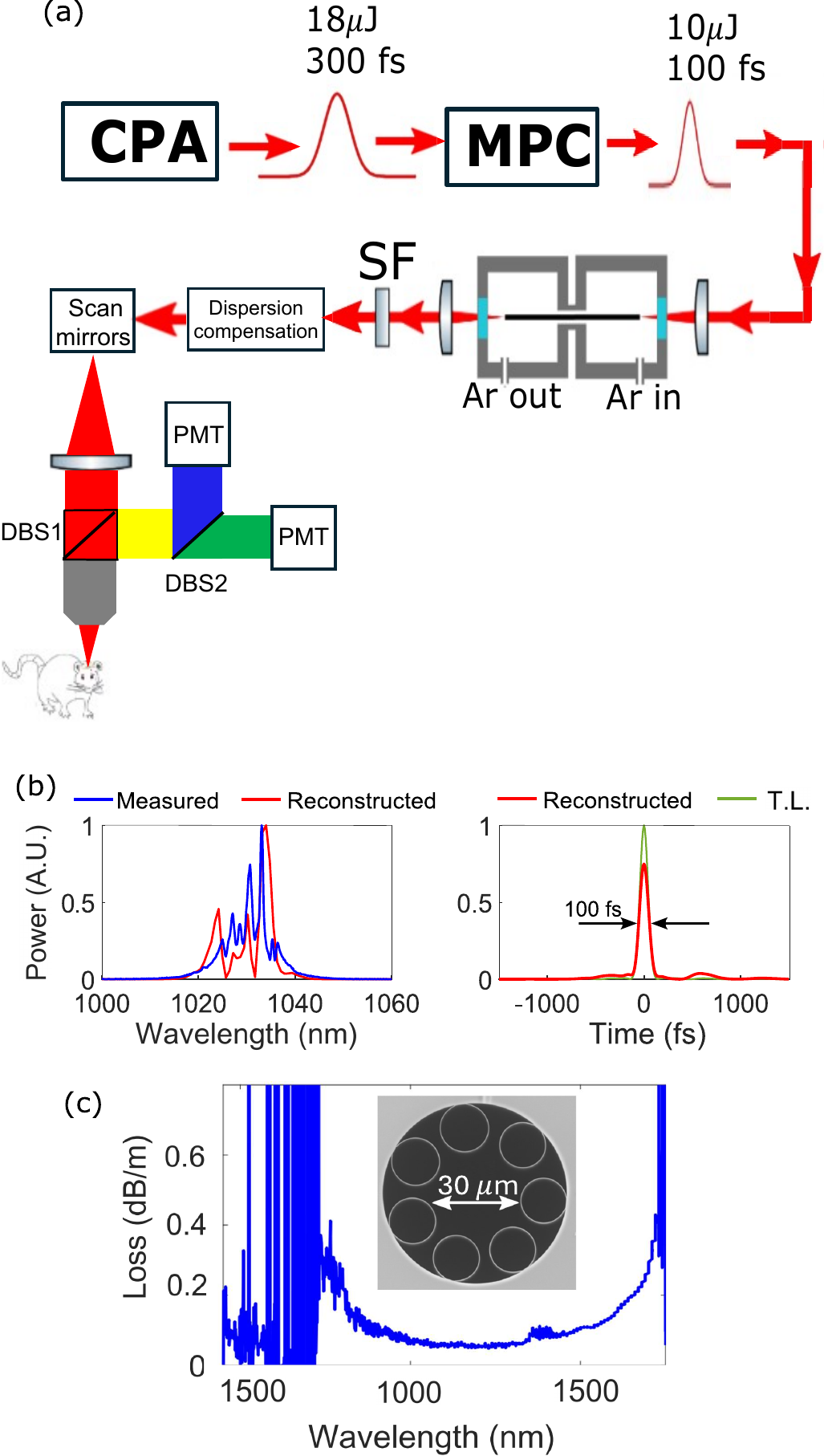}
\caption{(a) Schematic of the experimental system. The CPA emits 300-fs, 18-\si{\micro\joule} pulses, which are compressed to 100~fs in a multi-plate compressor (MPC). These pulses are coupled into an anti-resonant hollow-core fiber held in a pair of gas cells filled with Ar. The light that exits the fiber goes through a spectral filter (SF) to select the desired lobe. The filtered light then goes through a dispersive delay line to pre-compensate the dispersion of the microscope, and is aligned into the microscope. It is reflected off scanning mirrors and expanded, then collimated and sent through a dichroic beam splitter (DBS1) to the objective lens and onto the sample. Third-harmonic and three-photon fluorescence are reflected back through the objective lens and reflected of DBS1. DBS2, a second dichroic beamsplitter, spearates the third harmonic from the three-photon fluorescence, and each is sent to a photomultiplier tube (PMT). (b) Left: the measured and reconstructed spectrum of the pulse after the MPC. Right: the reconstructed temporal profile of the pulse after the MPC, overlaid with the transform limit of the reconstructed (red) spectrum in (b). The full-width at half-maximum (FWHM) of the measured intensity profile is marked by arrows.(c) The transmittance of the anti-resonant fiber. Inset: cross-sectional image of the same fiber.}
\label{fig: schematic and ARHCF}
\end{figure}
Because we desire strong nonlinear interaction with modest pulse energy, an anti-resonant hollow-core fiber (ARHCF) with a relatively-small core diameter of 30~\si{\micro\m} was chosen to guide the light while it interacts with the gas. This is the same fiber used in a prior work \cite{chen2022efficient_SSFS}. It allows single-mode (with very low overlap between the light and the glass structures), low-loss propagation at wavelengths between 800~nm and 1700~nm (Figure \ref{fig: schematic and ARHCF}(c)) and is held in a pair of gas cells (Figure \ref{fig: schematic and ARHCF}(a)). Ar was chosen as the gas because its response is purely electronic and it is readily available.

When evacuated, the ARHCF has a zero-dispersion wavelength near the resonance at 700~nm and anomalous dispersion at longer wavelengths; when filled with a normally-dispersive gas--as Ar is at the studied wavelengths--the zero-dispersion wavelength can be shifted to longer wavelengths. At the gas pressures used here, the waveguide still has anomalous dispersion at 1030~nm, which can create solitons in conjunction with the Kerr nonlinearity; however, it is generally desirable to avoid soliton formation in SPM-based spectral broadening. Operation with short segments of fiber and high gas pressures helps to avoid soliton formation: the dispersion is closer to zero, and the length of fiber may be  chosen so short that the field is not able to evolve to a soliton solution. Our current experimental setup cannot support a fiber shorter than 24 cm, so this length was used in the experiments.

An all-normal dispersion oscillator with a repetition rate of 31 MHz provides the initial pulses. These are stretched to a few hundred picoseconds duration in a chirped fiber Bragg grating and the repetition rate is reduced to 200 kHz. The pulses are amplified in three stages, the last being a rod-type single-mode waveguide \cite{limpert2006extended}, and are then dechirped in a volume Bragg grating to yield 18-\si{\micro\joule} and 300-fs pulses. 

\begin{figure}[h]
\hspace*{-0.5cm}  
\centering\includegraphics[width=0.5\textwidth]{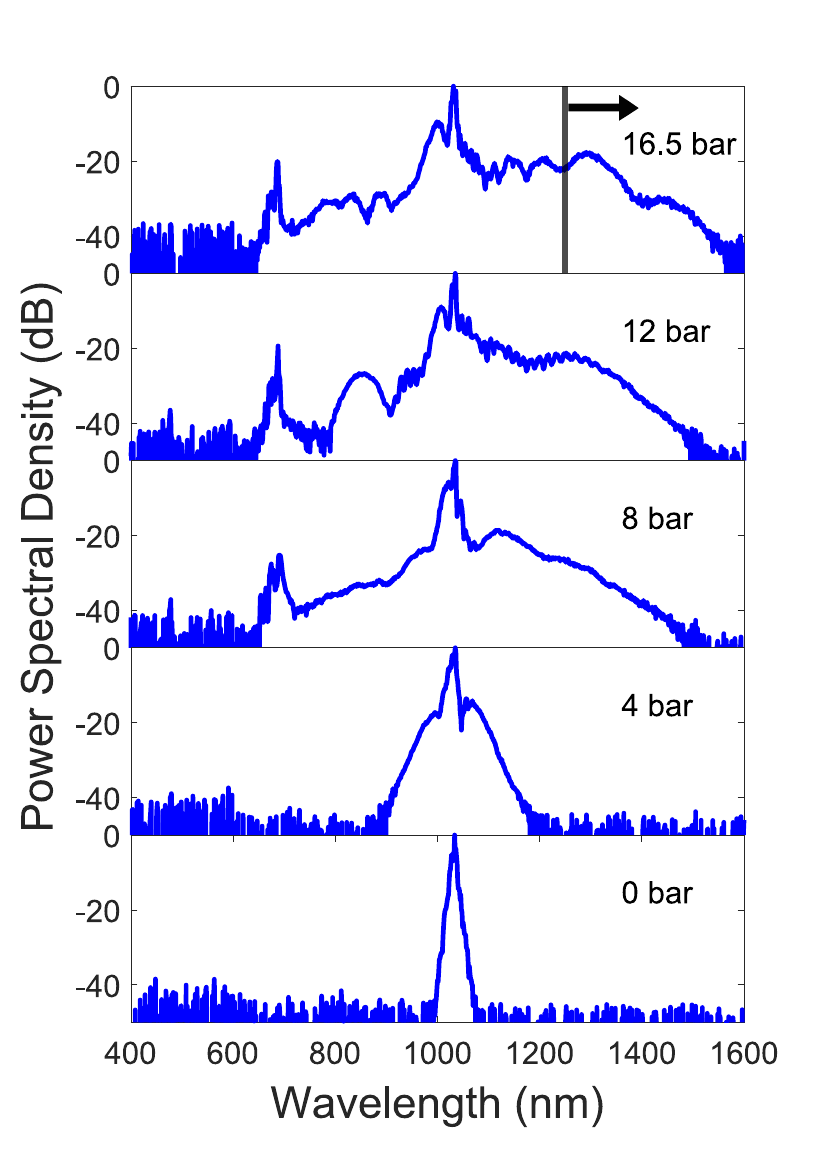}
\caption{Output spectra from the argon-filled fiber at the indicated pressures. The black vertical line in the top plot is the spectral edge of the long-pass filter. These plots understate the relative energy at 1300~nm due to the spectrometer's calibration.}
\label{fig:Spectra with increasing pressure}
\end{figure}
The effects of self-phase modulation strongly depend on the duration of the input pulse. We find, both in simulations and experiments, that SPM broadening of few-microjoule, 300-fs pulses does not generate useful light at the desired wavelengths if the argon pressure is below~100 bar. However, pulses about 100~fs long or shorter do yield the needed spectral broadening. Thus, a multi-plate compressor (MPC)\cite{vlasov2012MPC_theory, zhu2022_MPC_experiment} was used to compress the pulses from the amplifier. The MPC consists of 12 thin N-SF11 plates and a dispersive delay line. The beam from the CPA is focused onto the N-SF11 plates, where it spectrally broadens. The dispersive delay then removes the slight chirp induced on the pulse. The 300-fs pulses from the fiber amplifier can thus be compressed to between 60~fs and 100~fs. Pulses produced by the MPC had an energy of 10~\si{\micro\joule} and linear polarization. Around 7~\si{\micro\joule} was coupled into the ARHCF, which was filled with various pressures of Ar. The output from the ARHCF was filtered spectrally and the temporal profile of the filtered light was measured by frequency-resolved optical gating (FROG) for the pulses centered at wavelengths less than 1700~nm, and with a home-built interferometric autocorrelator for pulses at 1700~nm.

Pulses from the ARHCF were directed to a custom-built three-photon fluorescence microscope similar to the ones used in previous work  \cite{Shitong1, Shitong2, Shitong3}. A pair of prisms was used to pre-compensate the dispersion of the microscope objective, and a knife edge was placed between the prisms to remove spectral components above 1400~nm (Figure S12 in the Supplement), which resulted in a typical post-objective energy of 240 nJ. With an excitation NA of 0.75, the microscope maintains sub-cellular resolution across a field-of-view of 430 x 430~\si{\micro\meter^2} (Figure S12(c-e) in the Supplement).

\section{Results}
 With increasing Ar pressure, the output spectrum broadens (Figure \ref{fig:Spectra with increasing pressure}), as one would expect theoretically \cite{shi2023theoreticalSESS_gas}. Numerical simulations of this process with our experimental parameters are presented in the Supplemental Document. We will describe the generation and characterization of the pulses at 1300~nm in some detail. The spectrum broadens with increasing pressure until there is a lobe centered at 1300~nm. 
\begin{figure}[h!]
\hspace*{-0cm}  
\centering\includegraphics[width=0.5\textwidth]{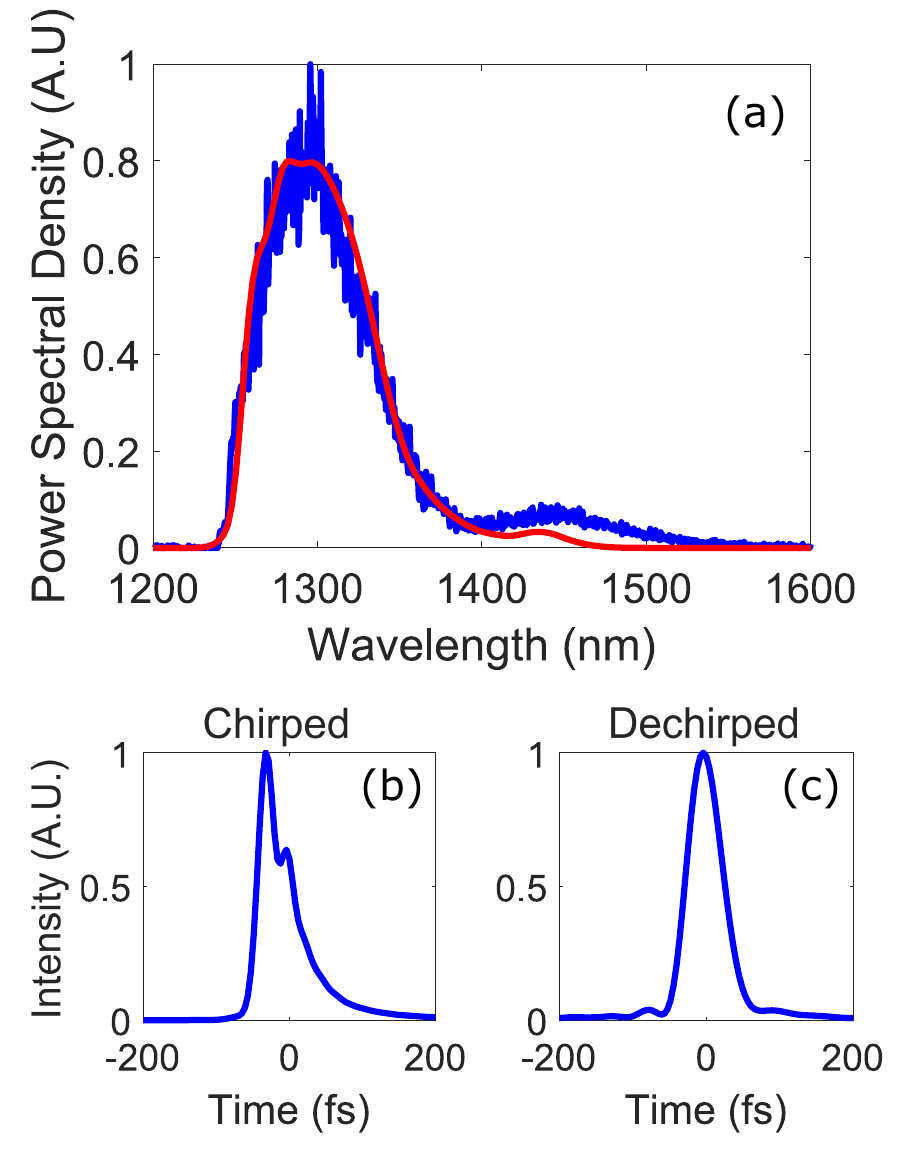}
\caption{(a) Blue: spectral profile of filtered spectral lobe; red: spectrum as reconstructed by FROG algorithm for dechirped pulse. (b) Temporal profile of the filtered lobe without dispersion compensation, and (c) temporal profile of the filtered lobe after dispersion compensation}
\label{fig:lobe at 1300}
\end{figure}
\begin{figure}[ht]
\hspace*{-0.5cm}  
\centering\includegraphics[width=0.5\textwidth]{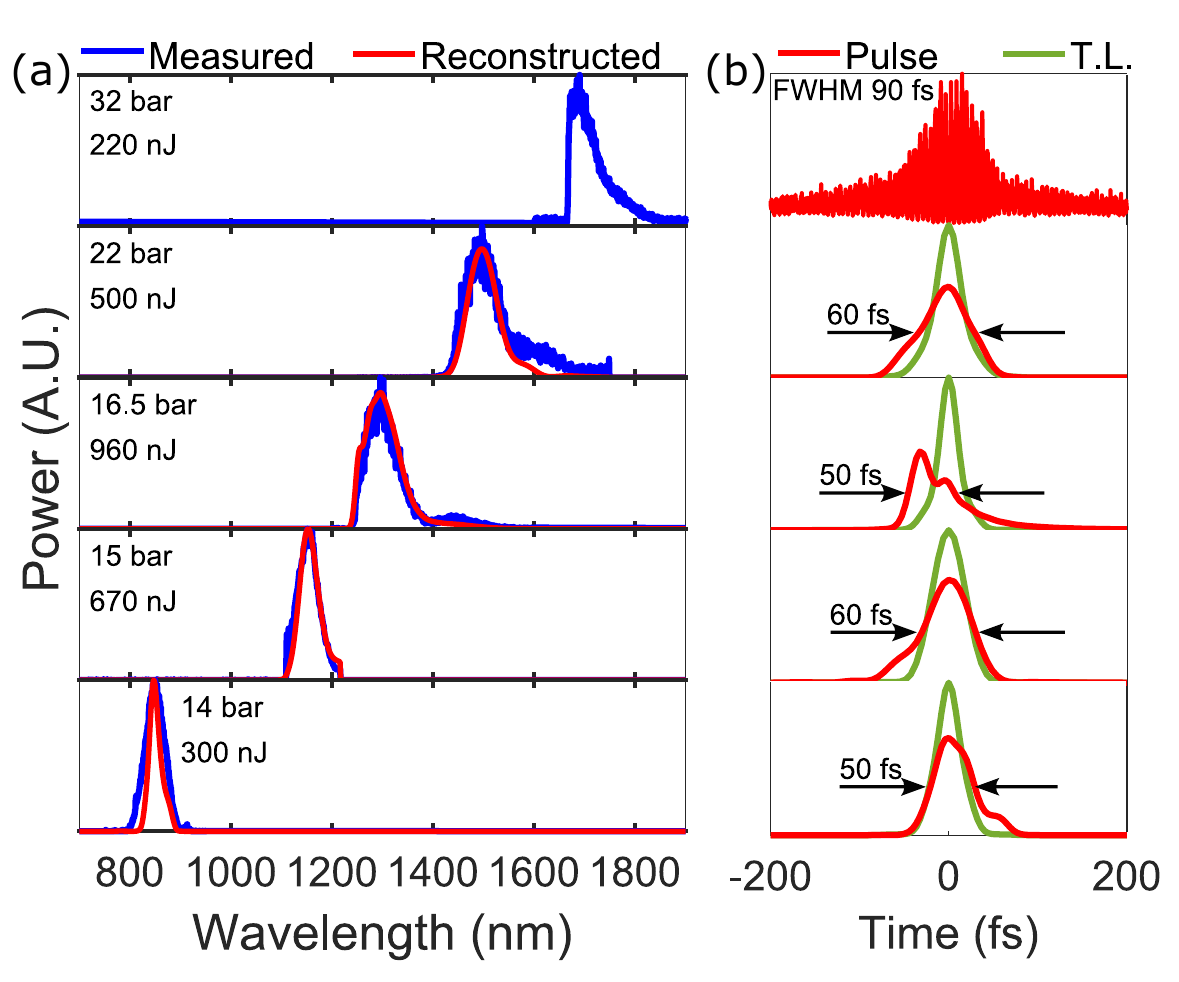}
\caption{Summary of tunable pulse-generation results. (a) Blue: spectral profiles generated at the indicated pressures. The energy in the spectral lobe depicted is also indicated. Red: the FROG reconstructions of the spectra of these spectral intervals. The data at 1300~nm, discussed above, are included here. (b) Red: the FROG reconstruction (or interferometric autocorrelation) of the pulse intensity as a function of time, corresponding to the reconstructed (red) spectra in (a). Green: transform-limited pulses corresponding to the FROG-reconstructed (red) spectra in (a). Each panel in (b) corresponds to the panel in (a) that is level with it. The FWHM durations of the reconstructed (red) pulses are marked by arrows. The top panel in (b), corresponding to a pulse centered at 1700~nm, is an interferometric autocorrelation, not a FROG reconstruction.  The horizontal axis for this panel has the same scale as the rest, but corresponds to delay rather than time. The pulse FWHM value of 90~fs in this panel assumes a deconvolution factor of 1.4.}
\label{fig: D and E}
\end{figure} 
 \begin{figure}[h!]  
\centering\includegraphics[width=.98\textwidth]{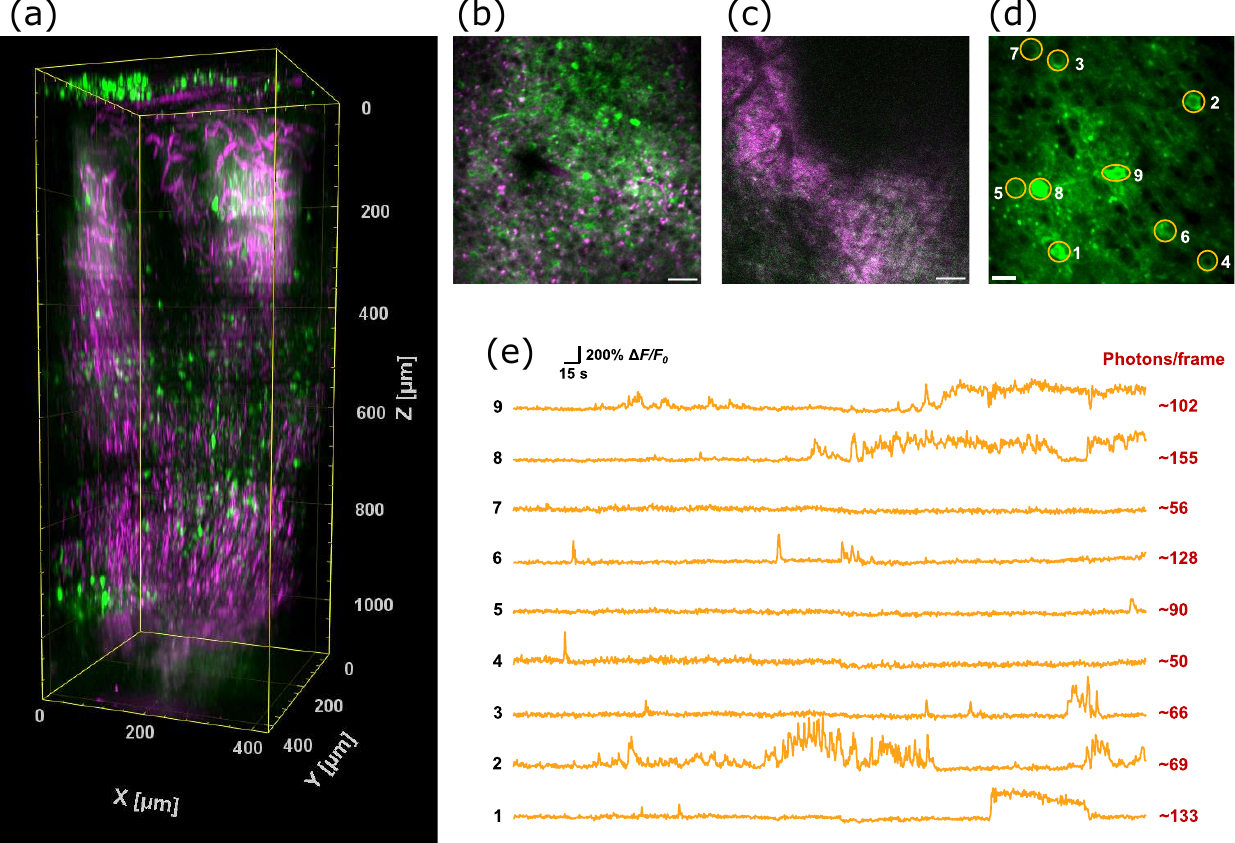}
\caption{The 1300-nm source enables in-vivo three-photon imaging deep (approximately 1.1~mm) into a transgenic GCaMP6s mouse brain. (a) 3D reconstruction of three-photon microscopy images (430~\si{\micro\meter} x 430~\si{\micro\meter} x 1110~\si{\micro\meter}) of the mouse brain (green, fluorescence; magenta, THG). A video of the stack shown here is available in the Supplementary Material.(b-c) Selected 2D frames at (b) 790-\si{\micro\meter} and (c) 1010~\si{\micro\meter} deep in a. Scale bars, 50~\si{\micro\meter}. (d) Neuronal activity recording of 9 neurons located at 932~\si{\micro\meter} beneath the dura. Imaging FOV, 215 x 215~\si{\micro\meter^2}; frame rate, 2.74 Hz; excitation power, 16~mW; laser repetition rate, 200~kHz; scale bar, 20~\si{\micro\meter}. (e) Spontaneous activity traces of the 9 individual neurons labeled in (d), with collected photons/neuron/frame written on the right. Recording length, 10 minutes. A video of the recorded activity shown in (d) and (e) is available in the Supplementary Material.}
\label{fig: imaging}
\end{figure}

 The spectral shape and temporal profile of this lobe were optimized by varying the input-pulse duration between 60 and 100~fs, with the best result obtained with 100-fs input pulses. Using a long-pass filter, 960~nJ were isolated at wavelengths longer than 1240~nm (Figure \ref{fig:lobe at 1300}(a)). The pulse in this spectral lobe, whose temporal profile is shown in Figure \ref{fig:lobe at 1300}(b), was found to have a full-width at half-maximum (FWHM) duration of around 50~fs, with a tail extending out to about 200~fs. Although the peak power is greater than 10~MW, we removed the tail with grating-pair dispersion compensation, as shown in Figure \ref{fig:lobe at 1300}(c). The FROG reconstructions for both the chirped and dechirped pulses have errors below 0.5\%. The agreement of the measured and reconstructed spectra (Figure \ref{fig:lobe at 1300}(a)) further supports the FROG reconstruction. The light in this spectral interval has a polarization extinction ratio of 13~dB.

For each wavelength of interest, the Ar pressure was varied, but the input pulse duration was fixed at 100~fs. The results are summarized in Figure \ref{fig: D and E}, and presented in greater detail in the Supplement. These results demonstrate that the source can provide multi-megawatt pulses throughout the octave spanning from 850~nm to 1700~nm, even without any dispersion compensation. The 100-fs pulses were used because they yield good performance at 1300~nm. By tailoring the duration of the pulses from the MPC as well as the gas pressure to produce the wavelength of interest, it should be possible to improve on the results presented here.

We performed both structural and functional three-photon in vivo imaging of neurons in transgenic GCaMP6s mice (CamKII-tTA/tetO-GCaMP6s). The animal procedures followed those described in Ref. \cite{Shitong3}. Briefly, we performed a craniotomy and placed a glass window to provide optical access to the brain (see Section 7 of the Supplement for more details). As Figure S12(a) in the Supplement shows, pulses of 60 fs duration can be obtained after the objective. The maximum pulse energy after the objective was 240 nJ, for an average power of 48 mW. We performed structural imaging of the entire cortical column and down to the external capsule, which extended from approximately 950 \si{\micro\meter} to 1100 \si{\micro\meter} depth (Figure \ref{fig: imaging}(a)-(c)). We consistently detected spontaneous neuronal activity more than 800~\si{\micro\meter} below the dura with sufficiently high numbers of photons/neuron/second for high-fidelity calcium transient recording \cite{Shitong3, Shitong4} (Figure \ref{fig: imaging}(d)-(e), Figure S13 in the Supplement). For functional imaging of the deepest active neurons we detected, at 930 ~\si{\micro\meter} below the dura, only about 50\% of the available laser power was used, which provides ample margin for use with microscopes with lower transmission at 1300 nm. 

To verify that the 1300-nm source provides similar excitation efficiency to existing sources for three-photon microscopy, we compare our calcium imaging results to those acquired with a NOPA in \cite{Shitong3}. The imaging experiments reported here were performed with approximately the same parameters that affect the signal (i.e., pulse duration, microscope collection efficiency, fluorophore concentration, three-photon action cross section, excitation wavelength, and excitation NA). Wang et al. reported that a pulse energy of 1.86 ± 0.27 nJ at the focus is required to detect 0.1 emission photon per excitation pulse. Assuming an effective attenuation length of 300 µm, our imaging results indicate that a pulse energy of 2.05 ± 0.26~nJ at the focus is required to achieve the same signal strength with the new source. The small difference could be due to variations in tissue properties such as scattering lengths and wavefront aberration. Therefore, within the uncertainties introduced by biological sample variations, the new source achieves the same excitation efficiency as the NOPA for three-photon microscopy.

\section{Discussion}  

The conversion efficiency from 1030~nm to 1300~nm within the ARHCF is almost 14\% with respect to the coupled pulse energy. In these proof-of-concept experiments, the coupling efficiency into the ARHCF was only 70\%, and the MPC efficiency was only 56\%.  As a result, the total efficiency of the system from the CPA output pulse to the wavelength-converted output was only about 5\%. The low efficiency of the MPC is primarily due to the use of a grating pair for the dispersive delay. The use of chirped mirrors rather than a grating pair should improve the MPC efficiency to 90\% or higher. Some improvement in the coupling into the ARHCF can also be expected.  With these changes, it should be quite possible to reach an overall efficiency of 10\%. Thus, a CPA that supplies 10-\si{\micro\joule} pulses  should be sufficient to achieve the performance described in Figure \ref{fig: D and E}. For use in high-speed imaging it will be important to increase the repetition rate of the source into the several-megahertz range, which should be possible. Short-pulse amplifiers with multi-megahertz repetition rates and above 20~W average power have been demonstrated \cite{kobayashi201310, aeroPULSE_FS50}, and some of the thermal effects that can limit Raman-based wavelength conversion \cite{Chini} will not occur in atomic gases. Overall, the high efficiency of our approach will ultimately reduce cost and enhance robustness by allowing the use of a lower power pump laser, and enable a high repetition-rate (>5 MHz) and high pulse-energy (>1 \si{\micro\joule}) source without increasing pump power beyond that of the existing NOPAs. Because optimum imaging requires matching the pulse energy and repetition rate with the imaging depth \cite{wang2020three}, such a source will deliver optimum imaging performance over a large range of depths by simply adjusting the repetition rate of the pump or attenuating the pulse energy. Currently, multiple NOPAs are required to achieve optimum conditions for imaging at depths between 600 \si{\micro\meter} and 1000 \si{\micro\meter} in a mouse brain, for example. Finally, this approach will create new opportunities to implement novel imaging techniques such as adaptive excitation for deep and fast imaging \cite{AES}. 

Extension of this approach to a broader range of wavelengths would be difficult in the ARHCF used here because, as Figure \ref{fig: schematic and ARHCF}(c) shows, this fiber has high losses at wavelengths longer than 1700~nm and shorter than 800~nm due to resonances. Energy scaling without broader wavelength tunability should be possible with the ARHCF used here by simply reducing the gas pressure. The use of a capillary rather than an ARHCF may allow broader spectral tunability, but this would require either higher pump pulse energy or higher gas pressure than those used here, because the mode-field area in any currently-existing capillary is larger than that of our ARHCF. On the other hand, the use of a capillary will also enable energy scaling for the same reason.

\section{Conclusion}
We have demonstrated a wavelength-tunable source of 100-fs pulses with peak powers well over 1~MW between 850 and 1700~nm based on self-phase modulation in Ar-filled fiber. In addition to the new wavelengths that are generated, the amplifier and compressor provide 100-MW pulses at 1030~nm. Our approach provides a new route to the generation of energetic sub-100-fs pulses that are tunable across the most-important wavelength windows for deep-tissue multiphoton microscopy. Structural and functional three-photon fluorescence images of tissue deep in mouse brain illustrate the effectiveness of the source.

\begin{backmatter}
\bmsection{Funding} National Institutes of Health (R01EB033179, U01NS128660); Office of Naval Research (N00014-20-1-2789);Army Research Office, Joint Directed Energy Transitions Office grant W911NF2410008.

\bmsection{Acknowledgments} WW acknowledges a Postdoctoral Fellowship from Shanghai Jiao Tong University for partial support of his studies at Cornell University. 

\bmsection{Disclosures} The authors declare no conflicts of interest. 

\bmsection{Data availability} Data underlying the results presented in this paper are
not publicly available at this time but may be obtained from the authors upon
reasonable request.

\bmsection{Supplemental document}
See Supplement 1 for supporting content. 

\end{backmatter}

\pagebreak
\bibliography{sample}

\end{document}


\maketitle

\section{Introduction}

In this supplement, we first show the simulation results that motivated this work, then discuss temporal and spectral data for the pulses we measured at the various wavelengths, and present a peak-power test we conducted for the pulse generated at 1300 nm. We also show the characterization of the microscope, and neuronal activity data for neurons 824 \si{\micro\m} below the dura for a different mouse from the one described in the main text. Finally, the methods for mouse surgeries are summarized. 

\section{Simulations}
\begin{figure}[ht]
\centering
\includegraphics[width=.7\linewidth]{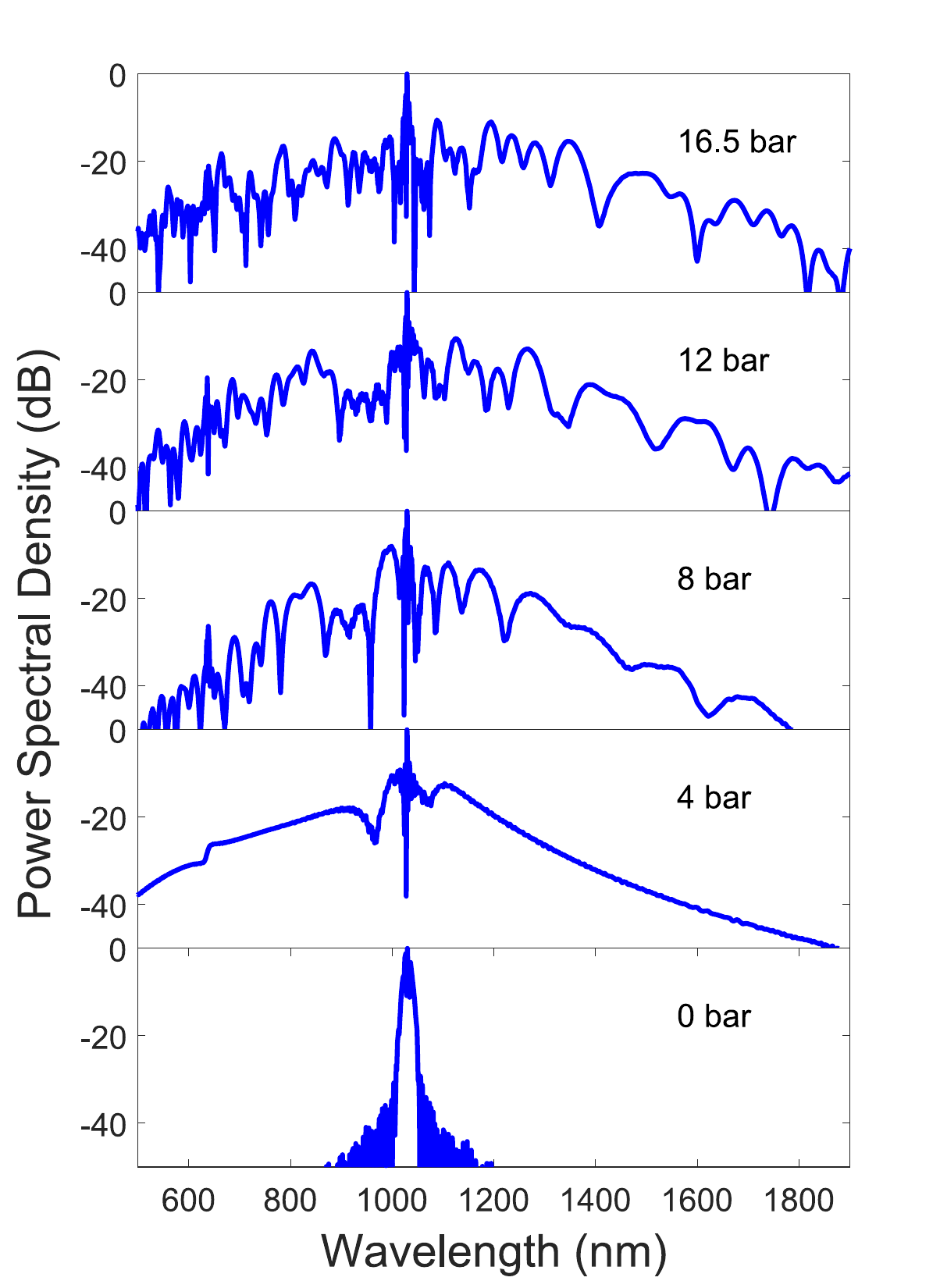}
\caption{Simulated spectra generated by launching pulses from MPC into 24 cm of ARHCF filled with Ar to indicated pressures.}
\label{fig:simulation trend}
\end{figure}

Although \cite{shi2023theoreticalSESS_gas} shows simulation results that demonstrate that SESS in atomic gas is a possible route to 1300 nm, we wanted to run simulations for our exact system: a compressed pulse undergoing SESS in argon gas, in the specific anti-resonant fiber used in the experiments. We used a fourth-order Runge-Kutta algorithm to simulate the propagation of our experimental 100-fs pulse (as reconstructed by FROG) by solving the unidirectional pulse propagation equation with dispersion, Kerr nonlinearity, and self-steepening. Figure \ref{fig:simulation trend} shows the trend, suggesting that this approach can be used to yield light at 1300 nm.
\pagebreak

\section{Discussion of results at wavelengths other than 1300 \MakeLowercase{nm}}
\subsection{850 nm}
\begin{figure}[ht]
\centering
\includegraphics[width=.7\linewidth]{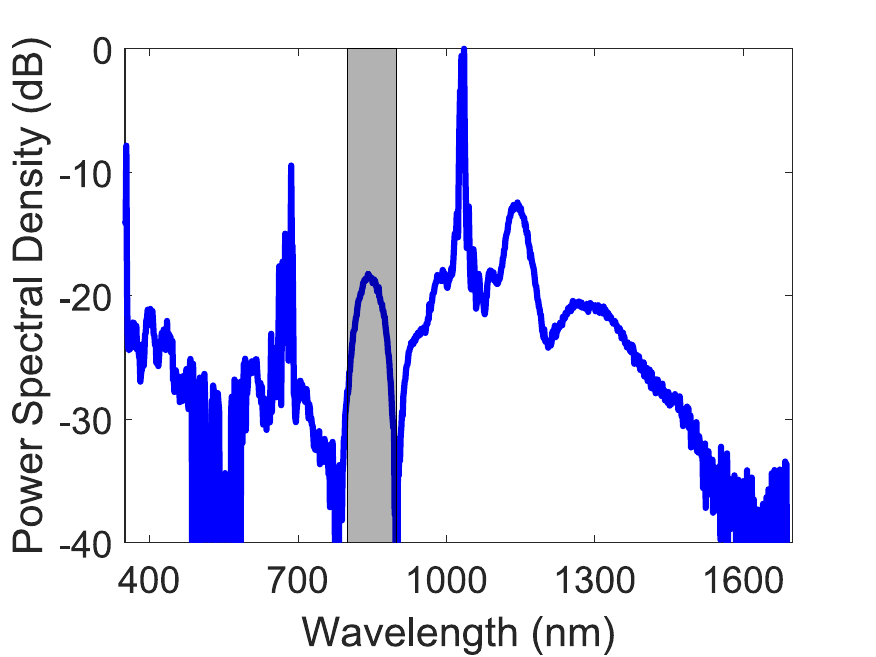}
\caption{The full spectrum at the output of the anti-resonant fiber pressurized with 14~bar of Ar. Shaded region: the lobe centered at 850 nm.}
\label{fig:full spectrum at 850 nm}
\end{figure}

\begin{figure}[ht]
\centering
\includegraphics[width=\linewidth]{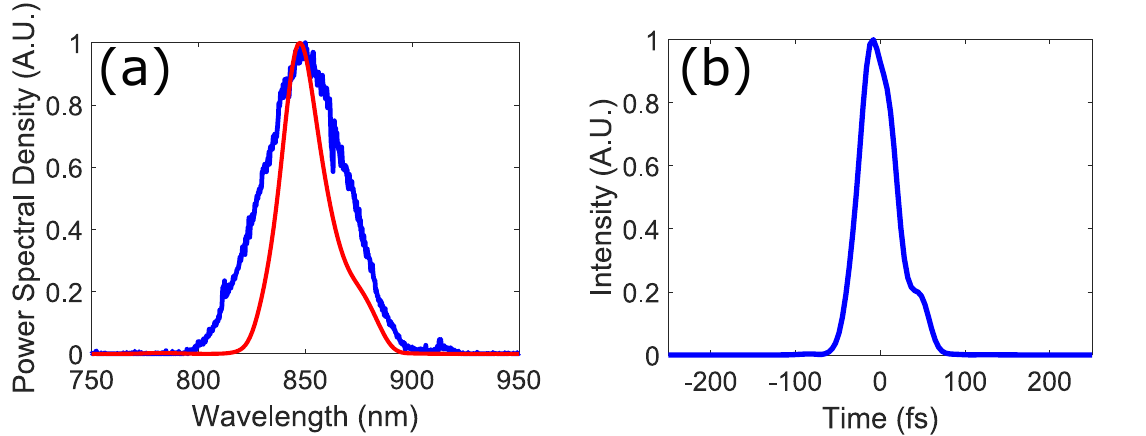}
\caption{(a) Blue: the filtered lobe centered at 850 nm, on linear scale. Red: the spectrum as reconstructed by FROG. (b) The corresponding pulse intensity, reconstructed by FROG.}
\label{fig:filtered spectrum and temporal profile at 850 nm}
\end{figure}

To generate a useful, high-power lobe at 850 nm, the fiber was filled with 14~bar of Ar. The output spectrum is shown on log scale in Figure \ref{fig:full spectrum at 850 nm}. Clearly, there is a lobe at 850 nm, as shown in the shaded region. We isolate this lobe with a pair of spectral filters as shown in Figure \ref{fig:full spectrum at 850 nm}, and show the spectrum in linear scale in Figure \ref{fig:filtered spectrum and temporal profile at 850 nm}(a), showing that the peak of the spectrum is at 850 nm. The light in this spectral band had an energy of 300~nJ (assuming that the output lens has an efficiency of 80\% at this wavelength, since it is coated to transmit longer wavelengths), and a polarization extinction ratio (PER) of 11.5~dB.


The temporal profile, as reconstructed by frequency-resolved optical gating (FROG) is shown in Figure \ref{fig:filtered spectrum and temporal profile at 850 nm}(b). The pulse has a full-width at half-maximum (FWHM) of 49~fs and the FROG error was below 1\%. It should be noted that this spectrum has the bandwidth for a 24-fs pulse, which our FROG was not able to reconstruct, due to the bandwidth limitations of FROG resulting from the finite phase-matching bandwidth of its crystal. The spectrum that the FROG did reconstruct has the bandwidth for 32-fs pulses, which is only slightly shorter than the pulse the FROG reconstructed. We hope to characterize this pulse more precisely in the future. In a worst-case scenario, the energy in the parts of the spectrum that were not reconstructed do not contribute at all to the peak power. In this case, we estimate that 60\% of the energy, or 180~nJ, are in the 49-fs pulse, making a peak power of around 3 MW. However, this worst-case scenario is highly unlikely, since the lobe is a clean SPM-lobe, and the part of the spectrum that was successfully reconstructed was near its transform-limit. As such, it is likelier that the entire lobe was closer to its transform limit of 24~fs. If it was roughly 30~fs long, which is probably the best-case scenario (due to the slight phase on the spectrum we did reconstruct), the peak power would be 10 MW.
\pagebreak
\subsection{1160 nm}
\begin{figure}[ht]
\centering
\includegraphics[width=.7\linewidth]{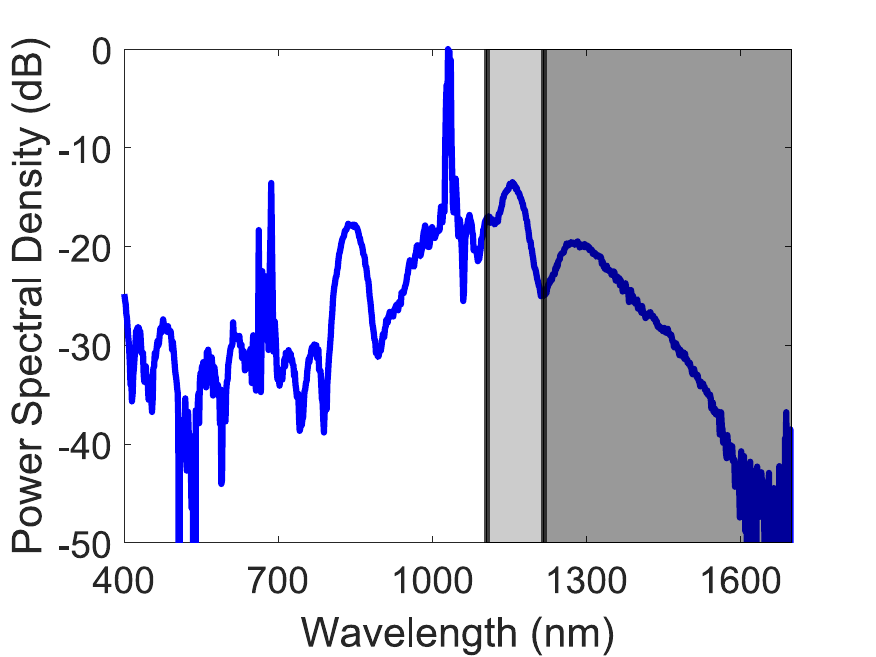}
\caption{The full spectrum at the output of the anti-resonant fiber pressurized with 15~bar of Ar. Shaded region: the interval isolated with a long-pass filter at 1106 nm. The darkly shaded region represents the part of the spectrum that does not interest us; the lightly shaded region is the lobe centered at 1160 nm.}
\label{fig:full spectrum at 1160 nm}
\end{figure}

The fiber was filled with 15~bar of Ar, and coupled the same MPC pulse through it. The full spectrum at the fiber output is shown in Figure \ref{fig:full spectrum at 1160 nm}. We isolated part of the spectrum with a long-pass filter at 1106 nm (the left vertical line in Figure \ref{fig:full spectrum at 1160 nm}. This part of the spectrum, which is represented by the entire shaded region in Figure \ref{fig:full spectrum at 1160 nm}, had an energy of 1.42 \si{\micro\joule} and PER of 11~dB. To study the lobe centered at 1160 nm, we inserted another long-pass filter at 1217 nm (the right vertical line in Figure \ref{fig:full spectrum at 1160 nm}) and found that the part of the pulse in this spectral region, which is represented by the darkly shaded region in Figure \ref{fig:full spectrum at 1160 nm}), had a total energy of 750~nJ. This means that the spectral lobe centered at 1160 nm (in the lightly shaded region in Figure \ref{fig:full spectrum at 1160 nm}) had an energy of 670~nJ. 

\begin{figure}[ht]
\centering
\includegraphics[width=.7\linewidth]{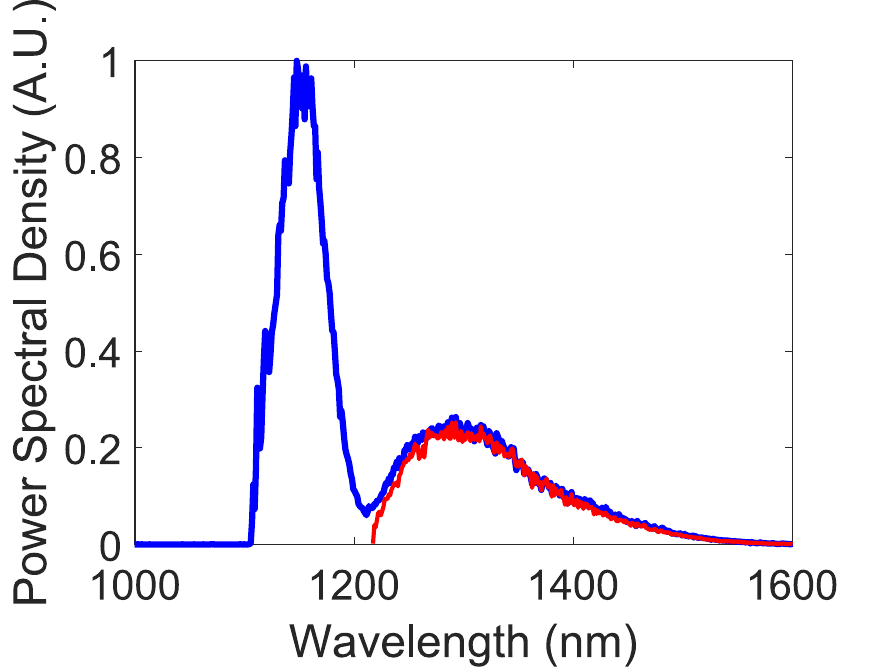}
\caption{The two isolated spectra from Figure \ref{fig:full spectrum at 1160 nm}), shown in linear scale. Blue: the spectrum filtered at 1106 nm, corresponding to the full shaded region in Figure \ref{fig:full spectrum at 1160 nm}). Red: the spectrum filtered at 1217 nm, corresponding to the darkly shaded region in Figure \ref{fig:full spectrum at 1160 nm}). We are not interested in the part of the spectrum represented by the red line.}
\label{fig:useful and garbage at 1160 nm}
\end{figure}

Figure \ref{fig:useful and garbage at 1160 nm} depicts both spectral intervals on linear scale. The spectrum in blue corresponds to the full, 1.42 \si{\micro\joule}-spectral interval beyond 1106 nm. The spectrum in red corresponds to the part of the spectrum beyond 1217 nm, which contains 750~nJ and does not interest us. As such, we claim 1420~nJ - 750~nJ = 670~nJ in the lobe centered at 1160 nm.

\begin{figure}[ht]
\centering
\includegraphics[width=\linewidth]{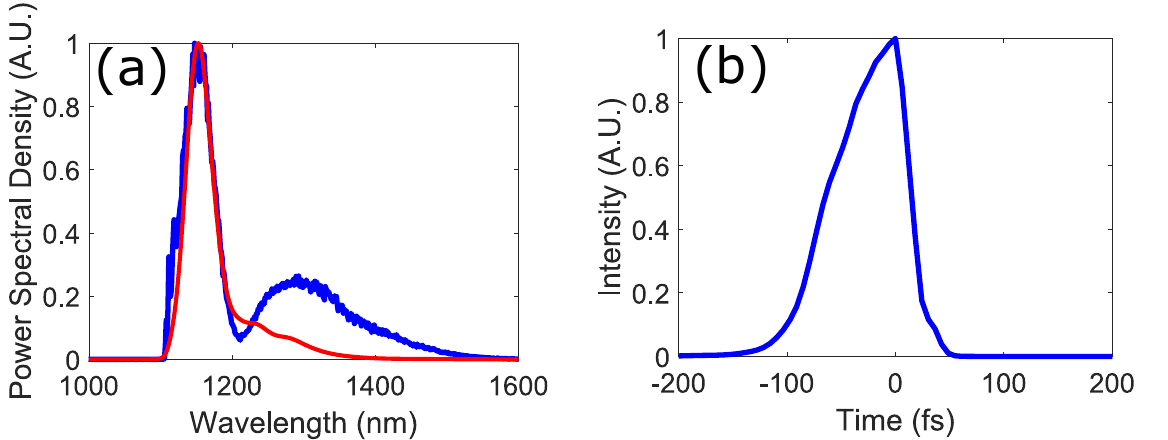}
\caption{(a) Blue: The full spectral interval beyond 1106 nm. Red: the spectrum reconstructed by the FROG. (b) The temporal profile reconstructed by the FROG, corresponding to the red spectrum in (a).}
\label{fig:FROG and spectra at 1160 nm.}
\end{figure}

Although we did not isolate the lobe at 1160 nm for spectral reconstruction due to a lack of a short-pass filter at 1217 nm, this did not end up mattering, since as mentioned above, the FROG itself can serve as a spectral filter. Here we were able to reconstruct the entire lobe that interested us, as in Figure \ref{fig:FROG and spectra at 1160 nm.}. Panel (a) depicts the full spectrum, which is the same as the blue line in Figure \ref{fig:useful and garbage at 1160 nm}. The red line in the same panel represents the spectrum as reconstructed by FROG, showing that FROG indeed acts as an effective spectral filter. The reconstructed pulse is shown in \ref{fig:FROG and spectra at 1160 nm.}(b). This measured pulse has a FWHM of 80~fs. The nominal transform-limited duration of this pulse is 40~fs. We did not attempt to dechirp it.

\begin{figure}[ht]
\centering
\includegraphics[width=\linewidth]{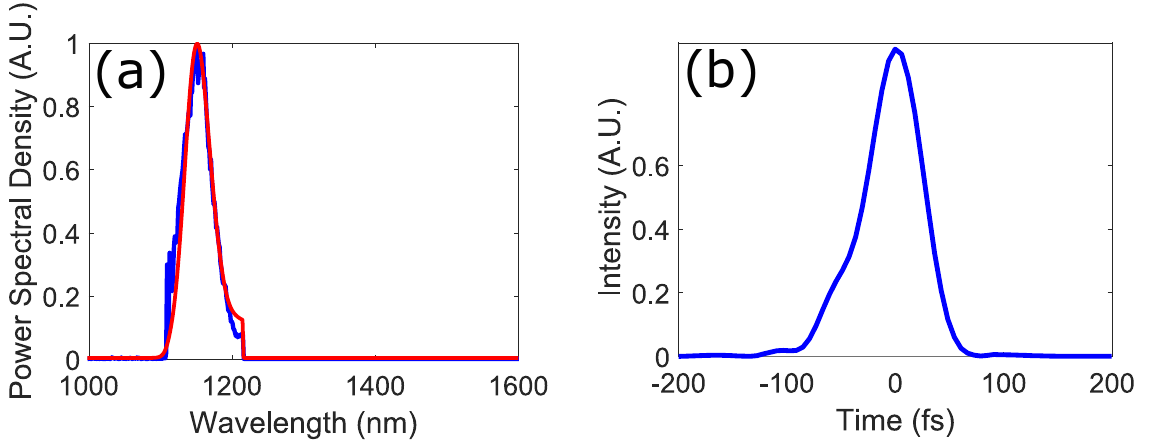}
\caption{(a) Blue: The full spectral interval beyond 1106 nm, numerically filtered at 1217 nm. Red: the spectrum reconstructed by the FROG, numerically filtered at 1217 nm. (b) The temporal profile reconstructed by the FROG, corresponding to the red spectrum in (a), with the spectral components of the pulse filtered, as in (a).}
\label{fig:FROG and spectra at 1160 filtered nm.}
\end{figure}

We note that the FROG reconstruction that yielded a FWHM of 80~fs assumes some energy in the part of the spectrum that is filtered out before we measured the energy of 670~nJ; Figure \ref{fig:FROG and spectra at 1160 nm.}(a) shows a tail in the reconstructed spectrum going all the way to 1400 nm. However, this energy is not in the 670-nJ lobe centered at 1160 nm, but rather in the 750-nJ part of the spectrum beyond 1217 nm. As such, we numerically inserted a short-pass filter at 1217 nm to see the temporal profile of the 670-nJ lobe that interests us. This shortens the FWHM of the reconstructed pulse to 60~fs, as shown in Figure 4 of the main text, and in Figure \ref{fig:FROG and spectra at 1160 filtered nm.} here.

\clearpage

\subsection{1500 nm}
\begin{figure}[ht]
\centering
\includegraphics[width=.7\linewidth]{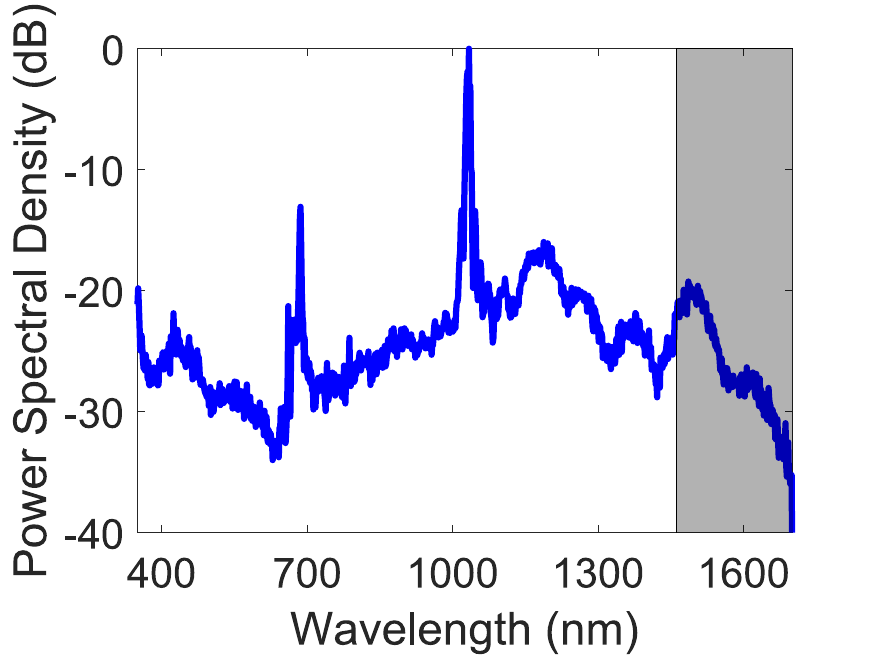}
\caption{The full spectrum at the output of the anti-resonant fiber pressurized with 22~bar of Ar. Shaded region: the lobe centered at 1500 nm, isolated with a long-pass filter at 1460 nm.}
\label{fig:full spectrum at 1500 nm}
\end{figure}

We pressurized the fiber with 22~bar of Ar, and coupled the same MPC pulse through it. The full spectrum at the fiber output is shown in Figure \ref{fig:full spectrum at 1500 nm}. We isolated part of the spectrum with a long-pass filter at 1460 nm (the shaded region in Figure \ref{fig:full spectrum at 1500 nm}. This part of the spectrum had an energy of 500~nJ and PER of 13~dB. 

\begin{figure}[ht]
\centering
\includegraphics[width=\linewidth]{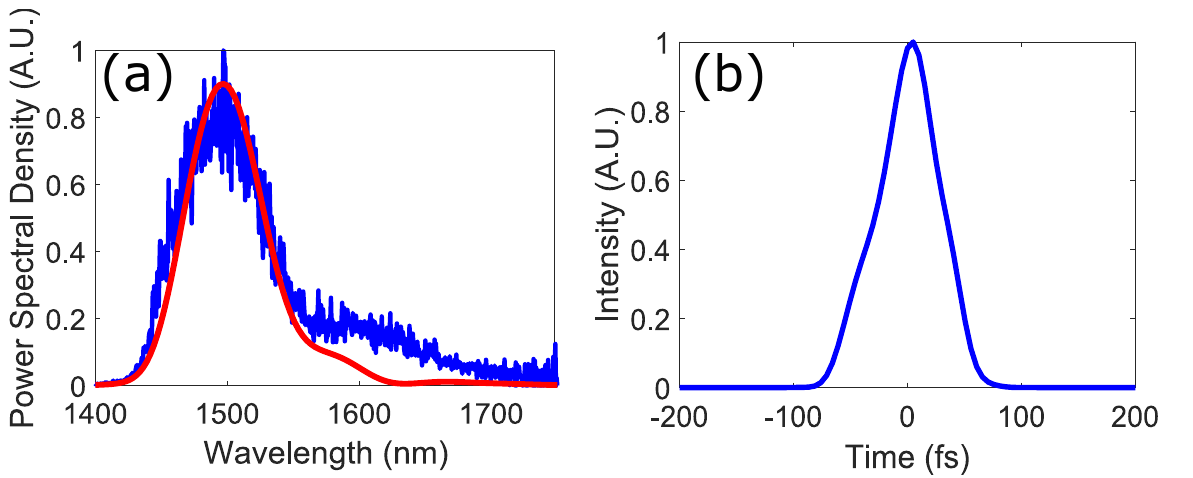}
\caption{(a) Blue: The full spectral region beyond 1460 nm, filtered with a long-pass filter (the shaded region in Figure \ref{fig:full spectrum at 1500 nm}), shown in linear scale. Red: FROG reconstruction of the spectrum of this spectral region. (b) FROG reconstruction of this lobe. }
\label{fig:lobe and FROG at 1500 nm}
\end{figure}

Figure \ref{fig:lobe and FROG at 1500 nm}(a) depicts the selected spectral interval in linear scale, and Figure \ref{fig:lobe and FROG at 1500 nm}(b) shows the FROG reconstruction of the temporal profile of the pulse. The agreement between the measured and retrieved FROG traces was within 1\% FROG error, and the reconstructed spectrum matched the measured spectrum well. The retrieved pulse has a duration of 62~fs, and a nominal transform-limited duration of 40~fs. We did not attempt to dechirp the pulse further.  
\pagebreak 
\subsection{1700 nm}
\begin{figure}[ht]
\centering
\includegraphics[width=.7\linewidth]{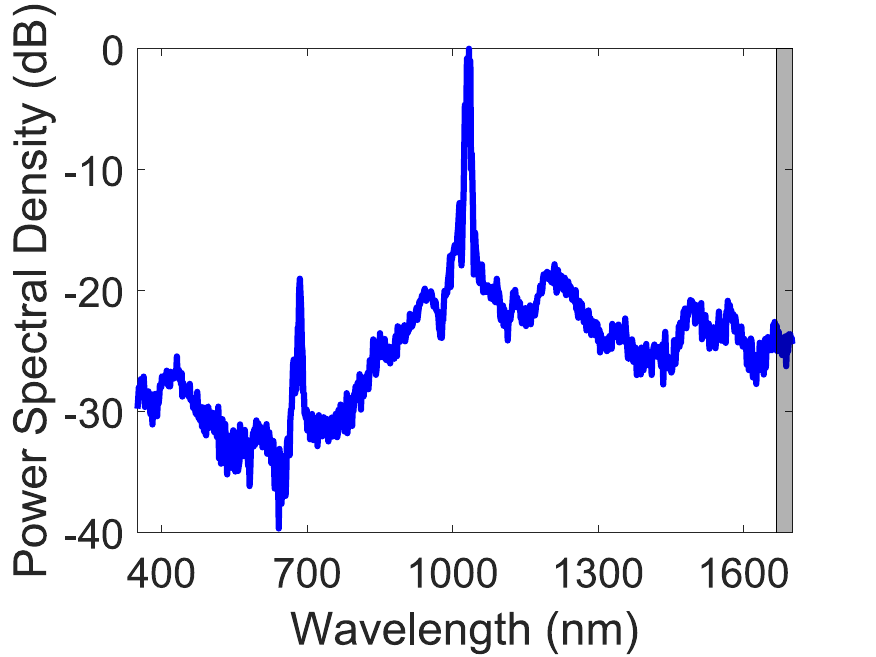}
\caption{The full spectrum at the output of the anti-resonant fiber pressurized with 32~bar of Ar. Shaded region: the lobe centered at 1700 nm, isolated with a long-pass filter at 1668 nm.}
\label{fig:full spectrum at 1700 nm}
\end{figure}

\begin{figure}[ht]
\centering
\includegraphics[width=\linewidth]{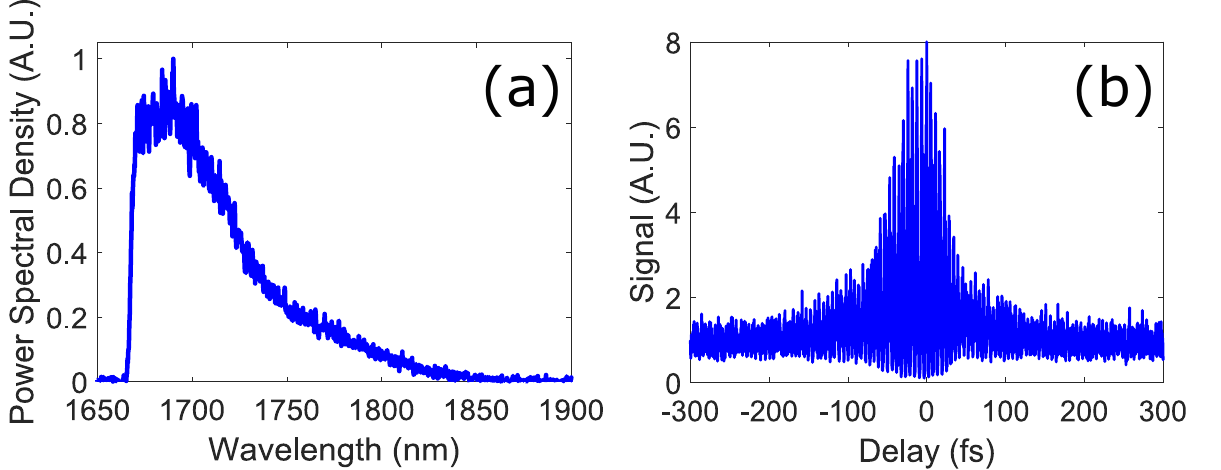}
\caption{(a)The spectral region filtered with a long-pass filter at 1668 nm (the shaded region in Figure \ref{fig:full spectrum at 1700 nm}), shown in linear scale. (b) Autocorrelation of this lobe, which corresponds to a Gaussian fit with a FWHM of 90~fs.}
\label{fig:lobe and FROG at 1700 nm}
\end{figure}

We pressurized the fiber with 32~bar of Ar. The full spectrum at the fiber output is shown in Figure \ref{fig:full spectrum at 1700 nm}. We isolated part of the spectrum with a long-pass filter at 1668 nm (the shaded region in Figure \ref{fig:full spectrum at 1700 nm}. The spectrometer does not reliably measure wavelengths far beyond 1700 nm, so we used a second spectrometer for the filtered lobe. This filtered lobe is depicted together with the corresponding autocorrelation trace in Figure \ref{fig:lobe and FROG at 1700 nm} (we measured an autocorrelation because our FROG device cannot measure these wavelengths).  This part of the spectrum had an energy of 220~nJ and PER of 11~dB. Assuming a Gaussian fit, the pulse should have a FWHM of around 90~fs. The transform-limited pulse has a FWHM of roughly 70~fs. 
\newpage
\clearpage
\section{Peak Power Test}
To provide further evidence that the pulse we generated at a carrier wavelength of 1300 nm is indeed a coherent pulse with high peak power, 60~nJ were split off from the chirped pulse and focused onto a 100-\si{\micro\m}-thick BBO crystal for second harmonic generation (SHG). The SHG conversion efficiency was compared with the theoretical value as follows.

If one assumes no pump depletion, no diffraction, and perfect phase matching for the entire spectrum, the energy efficiency of the SHG process will be
\begin{equation}
    \eta = \frac{\omega^2}{\epsilon_0 c^3}\bigg(\frac{d^2}{n^3}\bigg)\bigg( \frac{L^2}{A_{\text{eff}}} \bigg)\frac{\int (P(t))^2 dt}{\int (P(t)) dt}
    \label{Eq: SHG efficiency}
\end{equation}
Where $\omega$ is the central fundamental angular frequency, $\epsilon_0$ is the permittivity of vacuum; $c$ is the speed of light, $d$ is the element of the $\chi^{(2)}$ tensor relevant to this process; $n$ is the index of refraction of the crystal; $L$ is the length of the crystal, 100 \si{\micro\m} in our case; $A_{\text{eff}}$ is the effective area of the focused beam--approximately 50 \si{\micro\m^2}; and $P(t)$ is the pulse power as a function of time.

We used a value of 1.75 \si{\pm/\volt} for $d$\cite{alford2001wavelength}, and the ratio of integrals evaluates to 0.65 for our reconstructed pulse, after taking into account the dispersion of the beamsplitter. For these values, Equation \ref{Eq: SHG efficiency} predicts a SHG efficiency of roughly 50\%, whereas we achieved a conversion efficiency of 25\%. This puts a lower bound on the peak power of the chirped pulse as 70\% of the roughly 15 MW calculated by the FROG reconstruction. However, we believe that the peak power is higher, and that the discrepancy between theory and experiment can be easily explained by the fact that, as mentioned above, Equation \ref{Eq: SHG efficiency} neglects pump depletion, diffraction (which is significant over the length of the crystal, since the beam is tightly focused), and the finitude of the phase-matching bandwidth. 
\pagebreak
\section{Microscope Calibration}
\begin{figure}[ht]
\centering
\includegraphics[width=\linewidth]{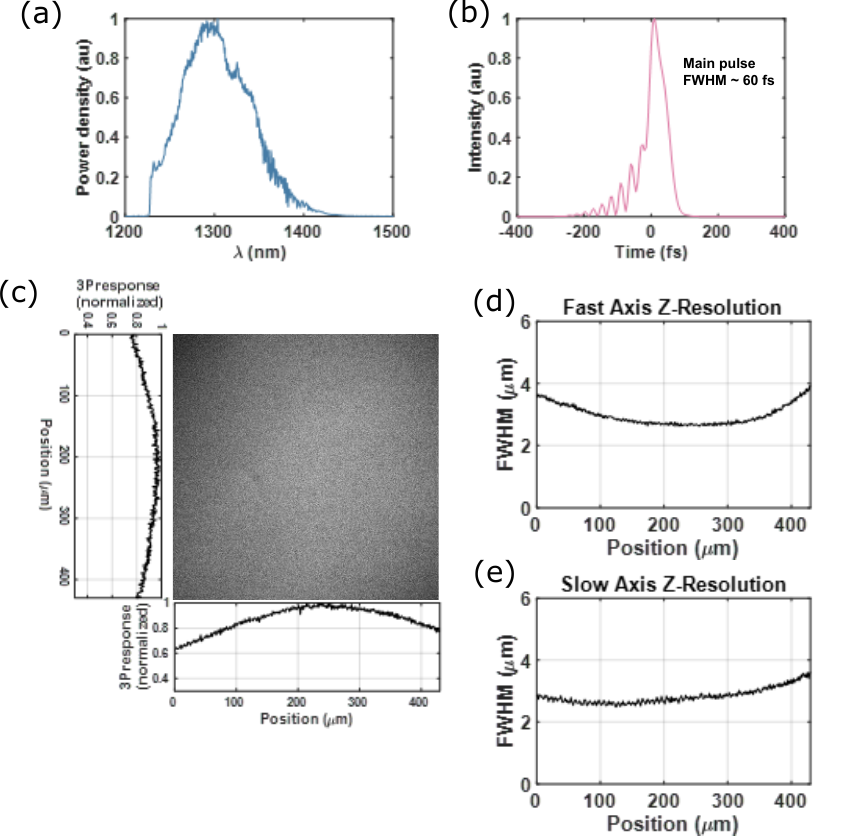}
\caption{Laser-microscope characterizations. (a) Measured spectrum of the laser into the microscope. (b) FROG-reconstructed temporal profile of the pulse under the objective lens, with dispersion pre-compensation. (c) Measured 3P response of the microscope using a fluorescein dye pool sample. (d-e) Measured axial resolution by volumetric scanning of a 500-nm thin fluorescent film of Rhodamine B dye. The FWHM of the axial intensity profiles is plotted along the (d) fast axis (horizontal direction in (c)) and the (e) slow axis (vertical direction in (c)). }
\label{fig:microscope calibration}
\end{figure}

Figure \ref{fig:microscope calibration} Shows the data mentioned briefly in the main text on the calibration of the microscope and laser system.
\pagebreak
\section{Neuronal Activity at 824 \si{\micro\m} Below the Dura in a Different Mouse (Mouse \#2)}
\begin{figure}[ht]
\centering
\includegraphics[width=\linewidth]{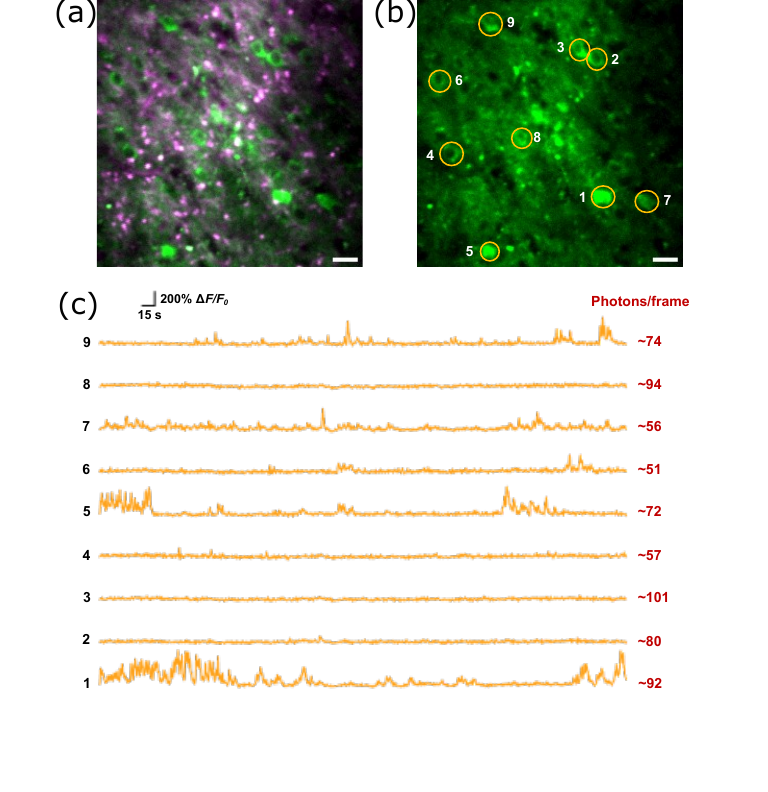}
\caption{Spontaneous neuronal activity recording of 9 neurons located at 824 \si{\micro\m} beneath the dura in a different transgenic GCaMP6s mouse brain (mouse \#2). (a) Average projection of a 10-min video recording (green, fluorescence; magenta, THG). (b) Fluorescence channel from a, with identified neurons labeled. Imaging FOV, 215 x 215 \si{\micro\m^2}; frame rate, 2.74 Hz; excitation laser power, 20~mW; laser repetition rate, 200 kHz; scale bars, 20 \si{\micro\m}. (c) Spontaneous activity traces of the 9 individual neurons labeled in b, with collected photons/neuron/frame written on the right. Recording length: 10 minutes.}
\label{fig:neuronal activity}
\end{figure}

Figure \ref{fig:neuronal activity} shows a recording of the activity of additional neurons, located at 824 \si{\micro\m} below the mouse's dura.

\section{Methods for Mouse Surgeries}
\subsection{Animals}
Transgenic mice (CamKII-tTA/tetO-GCaMP6s, n = 2, 2-8 months old, male) were used for all three-photon imaging experiments. Mice were housed in groups of 1–5 on a 12-h reverse light/ dark cycle before and after surgery. All animal experimentation and housing procedures were conducted in accordance with Cornell University Institutional Animal Care and Use Committee guidance.

\subsection{Chronic Craniotomy}
Prior to surgery, mice were given buprenorphine (intraperitoneally, 0.3 mg per kg of body weight) and dexamethasone (subcutaneously, 5 mg per kg of body weight). Animals were anesthetized with isoflurane (2 liters per minute of oxygen with 2-5\% isoflurane for induction, 1 liter of oxygen per minute with 1-3\% isoflurane during the surgery to maintain a constant breathing frequency). Body temperature was kept at 37.5 °C during the surgery with a standard small-animal heating pad (Kent Scientific). Eye ointment (Puralube veterinary ophthalmic ointment) was applied to protect the eyes during surgery. Animals were head-fixed in a stereotaxic apparatus (David Kopf Instruments). A 4-mm-diameter craniotomy was made over the right V1 (AP-2.7, ML-2.4 from bregma) with dura left intact. A glass window made of a circular 4-mm-diameter coverslip (4 mm Potomac, 170 \si{\micro\m}) glued to a circular glass ring with an inner diameter of 3.5 mm and outer diameter of 5.5 mm (3.5x5.5 mm Potomac, 170 \si{\micro\m}) was placed in the craniotomy, embedding the circular glass inside and keeping the glass ring on the skull around the craniotomy. The cranial window was sealed with cyanoacrylate superglue (Super Glue, ethyl cyanoacrylate) and with dental cement (C\&B Matabond). A custom-made titanium head-post was then attached to the skull with dental cement (C\&B Matabond). Ketoprofen (subcutaneously, 0.15 mg per kg of body weight) was administered immediately after surgery and then once daily for the following two consecutive days. Imaging was performed either on the same day of surgery, or at least 3 weeks after surgery when the tissue inflammation disappears, and mice were habituated to head fixation.

\pagebreak
\bibliography{sample}
